\documentclass[a4paper,UKenglish,cleveref, autoref, thm-restate,authorcolumns]{lipics-v2019}

\usepackage{graphicx}
\usepackage{xcolor}
\usepackage{xspace}
\usepackage{graphicx}
\usepackage[labelformat=simple]{subcaption}

\newcommand{\arity}{\textit{ar}}
\newcommand{\sub}{\textit{sub}}
\newcommand{\hs}{\textit{hs}}
\newcommand{\Var}{\textit{Var}}

\newcommand{\nf}{\textit{nf}\xspace}
\newcommand{\nfarray}{\texttt{nf}\xspace}
\newcommand{\nffreeze}{\texttt{nf\_read}\xspace}

\newcommand{\done}{\texttt{done}\xspace}
\newcommand{\garbagecollecting}{\texttt{garbage\_collecting}\xspace}
\newcommand{\freeindices}{\texttt{free\_indices}\xspace}
\newcommand{\nextfreebegin}{\texttt{next\_free\_begin}\xspace}
\newcommand{\nextfreeend}{\texttt{next\_free\_end}\xspace}
\newcommand{\nextfresh}{\texttt{next\_fresh}\xspace}
\newcommand{\refcounts}{\texttt{refcounts}\xspace}
\newcommand{\refcountsread}{\texttt{refcounts\_read}\xspace}
\newcommand{\heads}{\texttt{hss}\xspace}

\newcommand{\argu}{\texttt{arg}\xspace}
\newcommand{\maxarity}{\texttt{maxarity}\xspace}
\newcommand{\n}{\texttt{n}\xspace}

\newcommand{\rwrule}[1]{\textit{#1}}

\lstdefinelanguage{pseudo}{
  keywords = {procedure, for, from, to, do, if, then, return, true, false, while, cudaMemcpy, DeviceToDevice, DeviceToHost, HostToDevice, derive, rewrite, collect_free_indices, get_arg_index, switch, case, default, break, else, atomicSub, atomicInc, get_new_index, rewrite_Plus}
}

\title{Term Rewriting on GPUs} 

\author{Johri van Eerd}{Verum Software Tools BV, The Netherlands}{johri.van.eerd@verum.com}{}{}

\author{Jan Friso Groote}{Eindhoven University of Technology, The Netherlands \and \url{http://www.win.tue.nl/~jfg} }{j.f.groote@tue.nl}{https://orcid.org/0000-0003-2196-6587}{}

\author{Pieter Hijma}{VU Amsterdam, The Netherlands \and \url{http://www.few.vu.nl/~hha800} }{pieter@cs.vu.nl}{}{}

\author{Jan Martens}{Eindhoven University of Technology, The Netherlands}{j.j.m.martens@tue.nl}{}{}

\author{Anton Wijs}{Eindhoven University of Technology, The Netherlands \and \url{http://www.win.tue.nl/~awijs}}{a.j.wijs@tue.nl}{https://orcid.org/0000-0002-2071-9624}{}

\authorrunning{J. van Eerd, J.F. Groote, P. Hijma, J.J.M. Martens and A.J. Wijs} 

\Copyright{Johri van Eerd, Jan Friso Groote, Pieter Hijma, Jan Martens and Anton Wijs} 

\ccsdesc[500]{Theory of computation~Shared memory algorithms} 
\ccsdesc[500]{Theory of computation~Rewrite systems}

\keywords{Term rewriting, GPU, programming, parallel computing}
\category{} 

\relatedversion{} 

\supplement{}

\acknowledgements{This work is carried out in the context of the NWO AVVA project 612.001751.}

\nolinenumbers 

\hideLIPIcs  

\EventEditors{John Q. Open and Joan R. Access}
\EventNoEds{2}
\EventLongTitle{42nd Conference on Very Important Topics (CVIT 2016)}
\EventShortTitle{CVIT 2016}
\EventAcronym{CVIT}
\EventYear{2016}
\EventDate{December 24--27, 2016}
\EventLocation{Little Whinging, United Kingdom}
\EventLogo{}
\SeriesVolume{42}
\ArticleNo{23}

\begin{document}

\maketitle

\begin{abstract}
We present a way to implement term rewriting on a GPU. We do this by letting the GPU repeatedly 
perform a massively parallel evaluation of all subterms.
We find that if the term rewrite systems exhibit sufficient internal parallelism, GPU rewriting substantially 
outperforms the CPU.  Since we expect that our implementation can be further optimized, 
and because in any case GPUs will become much more powerful in the future, this suggests that 
GPUs are an interesting platform for term rewriting. 
As term rewriting can be viewed as a universal programming language, this also opens a route towards
programming GPUs by term rewriting, especially for irregular computations. 
\end{abstract}


\section{Introduction}
\label{sec:intro}
Graphics Processing Units (GPUs) increase in computational power much faster than the classical CPUs.  
GPUs are optimized for the highly parallel and regular computations that occur in graphics processing, but they become more and more
interesting for general purpose computations (for instance, see~\cite{bionetworks,Davidson14,gpuexplore2}). It is not without reason that modern super computers have large banks of
graphical processors installed in them \cite{Heldens2020Landscape,DBLP:conf/supercomputer/2019w}. GPU designers realize this and make GPUs increasingly suitable for irregular computations. For instance, they have added improved caches and atomic operations.


This raises the question to what extent the GPU can be used for more irregular computational tasks. The main limitation is that a highly parallel algorithm is needed to fully utilize the power of the GPU. For irregular problems it is the programmer's task to recognize the regularities in problems over irregular data structures such as graphs. 



The evaluation of term rewriting systems (TRSs), is an irregular problem that is interesting for the formal methods community.  For example term rewriting increases the expressiveness of models in the area of model checking~\cite{mcrl2} and the performance of term rewriting is a long-standing and important objective~\cite{duran2019rewrite}.
We recall that a term rewriting system that enjoys the Church-Rosser property is parallel in nature, in a sense that rewriting can take place at any point in the system and the order in which it takes place does not influence the outcome. This suggests a very simple model for parallel evaluation. Every processor can independently
work on its own section of the system and do its evaluation there. In this paper, we investigate whether and under which conditions term rewriting systems can be evaluated effectively on GPUs. We experimented with different compilation schemes from rewrite systems to GPU code and present here one where all processors evaluate all subterms in parallel. 
This has as drawback that terms that cannot be evaluated still require processing time. 
Terms can become discarded when being evaluated, and therefore garbage collection is required. All processors are also involved in this.

An earlier approach to inherently evaluate a program in parallel was done in the eighties. The Church-Rosser property for pure functional programs sparked interest from researchers, and the availability of cheap microprocessors
made it possible to assemble multiple processors to work on the evaluation of one single functional program. 
Jones et al.\ proposed GRIP, a parallel reduction machine design to execute functional programs on multiple microprocessors 
that communicate using an on-chip bus \cite{grip}. At the same time Barendregt et al.\ proposed the Dutch Parallel Reduction 
Machine project, that follows a largely similar architecture of many microprocessors communicating over a shared memory bus \cite{dprm}. 
Although technically feasible, the impact of these projects was limited, as the number of available processors was too small
and the communication overhead too severe to become a serious contender of sequential programming. GPUs offer a different infrastructure, with in the order of a thousand fold more processors and highly integrated 
on chip communication. Therefore, GPUs are a new and possibly better candidate for parallel evaluation of TRSs. 

Besides their use in the formal methods community, a term rewriting system is also a simple, yet universal mechanism for computation~\cite{HandbookKlop03}. A question that follows is whether this model for computation can be used to express programs for GPUs more easily. 

Current approaches for GPU programming are to make a program at a highly abstract level and transform it in a stepwise fashion to an optimal GPU program~\cite{Hijma2015Stepwise}. Other approaches are to
extend languages with notation for array processing tasks that can be sparked off to the GPU. Examples in the functional
programming world are 
Accelerate \cite{accelerate}, an embedded array processing language for Haskell, and Futhark \cite{Henriksen:2017:FPF:3062341.3062354}, 
a data parallel language which generates code for CUDA. While Futhark and Accelerate make it easier to use the power of the GPU, both approaches are tailored to highly regular problems. Implementing irregular problems over more complicated data structures remains challenging and requires the programmer to translate the problem to the regular structures provided in the language as seen in, for example,~\cite{henriksen2018modular}. 

We designed experiments and compared GPU rewriting with CPU rewriting of the same terms. We find that our implementation 
manages to employ 80\% of the bandwidth of the GPU for random accesses. For rewriting, random accesses are the performance bottleneck, and therefore our
implementation uses the GPU quite well. For intrinsically parallel rewrite tasks, the GPU outperforms a 
CPU with up to a factor 10. The experiments also show that if the number of subterms that can be evaluated in parallel is reduced,
 rewriting slows down quite dramatically. This is due to the fact that individual GPU processors are much slower than a CPU 
processor and GPU cycles are spent on non-reducible terms.


This leads us to the following conclusion. Term rewriting on a GPU certainly
has potential. Although our implementation performs close to random access peak
bandwidth, this does not mean that performance cannot be improved.  It
does mean that future optimizations need to focus on increasing regularity in
the implementation, especially in memory access patterns, for example by
grouping together similar terms, or techniques such as kernel unrolling in
combination with organizing terms such that subterms are close to the parent
terms as proposed by Nasre et al.~\cite{datadriven}.  Furthermore, we expect
that GPUs quickly become faster, in particular for applications with random
accesses.

However, we also observe that when the degree of parallelism in a term is reduced, it is better to let the CPU
do the work. This calls for a hybrid approach where it is dynamically decided whether a term is to be evaluated on the 
CPU or on the GPU depending on the number of subterms that need to be rewritten. This is future work. We also
see that designing inherently parallel rewriting systems is an important skill that we must learn to master. 


Although much work lies ahead of us, we conclude that using GPUs to solve term rewriting processes is promising. It allows for abstract programming independent of the hardware details of GPUs, and it offers
the potential of evaluating appropriate rewrite systems at least one order, and in the future orders of magnitude faster than a CPU. 

Related to this work is the work of Nasre et al.~\cite{nasre2013morph} where parallel graph mutation and rewriting programs for both GPUs and CPUs are studied. In particular they study Delaunay mesh refinement(DMR) and points-to-analysis (PTA). PTA is related to term rewriting in a sense that nodes do simple rule based computations, but it is different in the sense that no new nodes are created. In DMR new nodes and edges are created but the calculations are done in a very different manner. The term rewriting in this work can be seen as a special case of graph rewriting, where every symbol is seen as a node and the subterms as the edges.   
%
%

\section{Preliminaries}
\label{sec:prelim}

We introduce term rewriting, what it means to apply rewrite rules, and an overview of the CUDA GPU computing model.

\paragraph*{Term rewrite systems}

A \emph{Term Rewrite System} (\emph{TRS}) is a set of rules. Each rule is a pair of terms, namely a left hand side and a right hand side. 
Given an arbitrary term $t$ and a TRS $R$, rewriting means to replace occurrences in $t$ of the left hand side of a rule in $R$ by the 
corresponding right hand side, and then repeating the process on the result. 
 
Terms are constructed from a set of variables $V$ and a set of function symbols $F$. A function symbol is applied to 
a predefined number of arguments or subterms. We refer to this number as the \emph{arity} of the function symbol, and denote the arity of a function symbol $f$ by $\arity(f)$. If $\arity(f) = 0$, we say $f$ is a \emph{constant}.
Together, the sets $V$ and $F$ constitute the \emph{signature} $\Sigma = (F, V)$ of a TRS. The set of terms $T_\Sigma$ over a signature $\Sigma$ is inductively defined as the smallest set satisfying:
\begin{itemize}
\item If $t \in V$, then $t \in T_\Sigma$;
\item If $f \in F$, and $t_i\in T_\Sigma$ for $1 \leq i \leq \arity(f)$), 
then $f(t_1, \ldots, t_{\arity(f)}) \in T_\Sigma$.
\end{itemize}
With $\sub_i(t)$, we refer to the $i$-th subterm of term $t$.
The \emph{head symbol} of a term $t$ is defined as $\hs(f(t_1, \ldots, t_k)) = f$. If $t \in V$, $\hs(t)$ is undefined.
With $\Var(t)$, we refer to the set of variables occurring in term $t$. It is defined as follows:
\[
\Var(t) =
\begin{cases}
\{ t \} & \textrm{if}\ t \in V,\\
\bigcup_{1 \leq i \leq \arity(t)} \Var(t_i) & \textrm{if}\ t = f(t_1, \ldots, t_{\arity(f)}).  \\
\end{cases}
\]

\begin{definition}[Term rewrite system]
\label{def:trs}
A TRS $R$ over a signature $\Sigma$ is a set of pairs of terms, i.e., $R \subseteq T_\Sigma \times T_\Sigma$. Each pair $(l,r) \in R$ is called a \emph{rule}, and is typically denoted by $l \rightarrow r$. Each rule $(l,r) \in R$ satisfies two properties:
(1) $l \not\in V$, and (2) $\Var(r) \subseteq \Var(l)$.
\end{definition}

Besides the two properties for each rule $(l,r) \in R$ stated in Definition~\ref{def:trs}, we assume that each variable $v \in V$ occurs at most once in $l$. A TRS with rules not satisfying this assumption can be rewritten to one that does not contain such rules. 
Given a rule $l \rightarrow r$, we refer to $l$ as the left-hand-side (LHS) and to $r$ as the right-hand-side (RHS).

\begin{definition}[substitution]
For a TRS $R$ over a signature $\Sigma = (F,V)$, a \emph{substitution} $\sigma: V \to T_\Sigma$ maps variables to terms. We write $t\sigma$ for a substitution $\sigma$ applied to a term $t \in T_\Sigma$, defined as $\sigma(t)$ if $t \in V$, and $f(t_1\sigma,\ldots,t_{\arity(f)}\sigma)$ if $t = f(t_1,\ldots,t_{\arity(f)})$.
\end{definition}

Substitutions allow for a \emph{match} between a term $t$ and rule $l \rightarrow r$. A rule $l \rightarrow r$ is said to match $t$ iff a substitution $\sigma$ exists such that $l\sigma = t$. If such a $\sigma$ exists, then we say that $t$ \emph{reduces} to $r\sigma$. 
A match $l\sigma$ of a rewrite rule $l \rightarrow r$ is also called a \emph{redex}.

%
%

A term $t$ is in \emph{normal form}, denoted by $\nf(t)$, iff its subterms are in normal form and there is no rule $(l, r) \in R$ and substitution $\sigma$ such that $t = l\sigma$.

As an example, Listing~\ref{list:trs} presents a simplified version of a merge sort rewrite system with an input tree of depth 2 consisting of empty lists (\rwrule{Nil}). After the \texttt{sort} keyword, a list is given of all function symbols. After the keyword \texttt{eqn}, rewrite rules are given in the form LHS = RHS. The set of variables is given as a list after the \texttt{var} keyword. The \texttt{input} section defines the input term. In this example, all rewrite rules for functions on (Peano) numbers and Booleans are omitted, such as the less than (\rwrule{Lt}) rule for natural numbers and the \rwrule{Even} and \rwrule{Odd} rules for lists, which create lists consisting of all elements at even and odd positions in the given list, respectively. The potential for parallel rewriting is implicit and can be seen, for instance, in the \rwrule{Sort2} rule. The two arguments of \rwrule{Merge} in the RHS of  \rwrule{Sort2} can be evaluated in parallel. Note that \rwrule{Nil()}, \rwrule{Zero()} and \rwrule{S(Nat)} are in normal form, but other terms may not be. The complete TRS is given in Appendix~\ref{appendix:multisort_trs}.

\begin{lstlisting}[caption={A TRS for merge sort in a binary tree of lists}, label={list:trs},captionpos=t,float=t,abovecaptionskip=-\medskipamount]
sort  List = Nil() | Cons(Nat, List) | Sort(List) | ...;
      Tree = Leaf(List) | Node(Tree,Tree);
      Nat = Zero() | S(Nat);
      
var X : Nat; Y : Nat; L : List; M : List;

eqn   Merge(Nil(), M) = M;
      Merge(L, Nil()) = L;
      Merge(Cons(X, L), Cons(Y, M)) = Merge2(Lt(X,Y), X, L, Y, M);

      Merge2(True(), X, L, Y, M) = Cons(X, Merge(L, Cons(Y, M)));
      Merge2(False(), X, L, Y, M) = Cons(Y, Merge(Cons(X, L), M));

      Sort(L) = Sort2(Gt(Len(L), S(Zero())), L);
      Sort2(False(), L) = L;
      Sort2(True(), L) = Merge(Sort(Even(L)), Sort(Odd(L)));

input Node(Leaf(Sort(...)),Leaf(Sort(...)));
\end{lstlisting}

A TRS is \emph{terminating} iff there are no infinite reductions possible. For instance, the rule $f(a) \rightarrow f(f(a))$ leads to an 
infinite reduction. In general, determining whether a given TRS is terminating is an undecidable problem \cite{termination}.

The computation of a term in a terminating TRS is the repeated application of rewrite rules until the term is in normal form. Such a computation is also called a \emph{derivation}. Note that the result of a derivation may be non-deterministically produced. Consider, for example, the rewrite rule $r = (f(f(x)) \rightarrow a)$ and the term $t = f(f(f(a)))$. Applying $r$ on $t$ may result in either the normal form $a$ or $f(a)$, depending on the chosen reduction. To make rewriting deterministic, a \emph{rewrite strategy} is needed. We focus on the \emph{inner-most} strategy, which gives priority to selecting redexes that do not contain other redexes. In the example, this means that the LHS of $r$ is matched on the inner $f(f(a))$ of $t$, leading to $f(a)$.

\begin{lstlisting}[language=pseudo,caption={A derivation procedure for term $t$, and a rewrite procedure for head symbol $f$},label=list:derive,captionpos=t,float=t,abovecaptionskip=-\medskipamount]
procedure derive($t, R$):
  while $\neg \nf(t)$ do
    for $i \in \{ 1, \ldots, \arity(t) \}$ do
      if $\neg \nf(\sub_i(t))$ then
        derive($\sub_i(t)$)
    $t \leftarrow$ rewrite$_{\hs(t)}$($t, R$)

procedure rewrite$_f$($t, R$):
  rewritten $\leftarrow$ false
  for $(l \rightarrow r) \in \{ (l, r) \in R \mid \hs(l) = f \}$ do
    if $\exists \sigma: V \to T_\Sigma. l\sigma = t$ then
      $t \leftarrow r\sigma$; rewritten $\leftarrow$ true; break
  if $\neg$rewritten then $\nf(t) \leftarrow$ true
  return $t$
\end{lstlisting}

Algorithmically, (inner-most) rewriting is typically performed using recursion.
Such an algorithm is presented in Listing~\ref{list:derive}.
As long as a term $t$ is not in normal form (line 2), it is first checked whether all its subterms are in normal form (lines 3-4). For each subterm not in normal form, \texttt{derive} is called recursively (line 5), by which the inner-most rewriting strategy is achieved. If the subterms are checked sequentially from left to right, we have \emph{left-most} inner-most rewriting. A parallel rewriter may check the subterms in parallel, since inner-most redexes do not contain other redexes. Once all subterms are in normal form, the procedure \texttt{rewrite}$_{\hs(t)}$ is called (line 6).

For each head symbol of the TRS, we have a dedicated rewrite procedure. The structure of these procedures is also given in Listing~\ref{list:derive}. The variable \texttt{rewritten} is used to keep track
of whether a rewrite step has been performed (line 9). For each rewrite rule $(l, r)$ with $\hs(l) = f$, it is checked whether a match between $l$ and $t$ exists, and if so, $l \rightarrow r$ is applied on $t$ (lines 10-12). If no rewrite rule was applicable, it is concluded that $t$ is in normal form (line 13).

\paragraph*{GPU basics}

In this paper, we focus on NVIDIA GPU architectures and the Compute Unified Device Architecture (CUDA) interface.
However, our algorithms can be straightforwardly applied to any GPU architecture with a high degree of hardware multithreading and the SIMT (Single Instruction Multiple Threads) model.

\begin{figure}[t]
\centering
\scalebox{0.65}{
\includegraphics{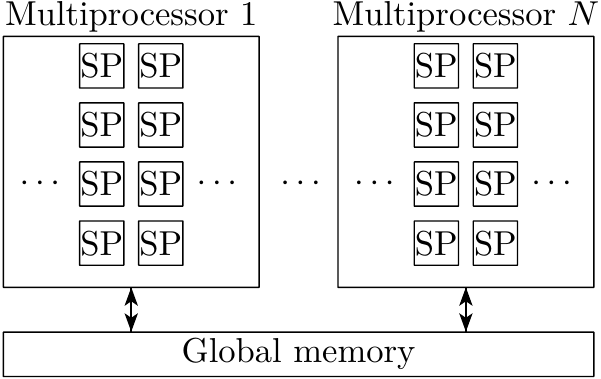}
}
\caption{Hardware model of CUDA GPUs}
\label{fig:cuda}
\end{figure}

CUDA is NVIDIA's interface to program GPUs. It extends the C++ programming language. CUDA includes special declarations to explicitly place variables in the various types of memory (see Figure~\ref{fig:cuda}), predefined keywords to refer to the IDs of individual threads and blocks of threads, synchronisation statements, a run time API for memory management, and statements to define and launch GPU functions, known as \emph{kernels}. In this section we give a brief overview of CUDA. More details can be found in, for instance,~\cite{GPU}.

A GPU contains a set of streaming multiprocessors (SMs), and each of those contains a set of streaming processors (SPs), see Figure~\ref{fig:cuda}. The NVIDIA \textsc{Turing Titan RTX}, which we used for our experiments, has 72 SMs, each having 64 SPs, which is in total 4608 SPs. 

A CUDA program consists of a {\em host} program running on the CPU and a collection of CUDA kernels. Kernels describe the parallel parts of the program and are launched from the host to be executed many times in parallel by different threads on the GPU.
It is required to specify the number of threads on a kernel launch and all
threads execute the same kernel. Conceptually, each thread is executed by an SP. In general, GPU threads are grouped in blocks of a predefined size, usually a power of two. A block of threads is assigned to a multiprocessor.

Threads have access to different kinds of memory. Each thread has a number of
on-chip registers to store thread-local data. It allow fast access. 
All the threads have access to the \emph{global memory} which is large
(on the \textsc{Titan RTX} it is 24 GB), but slow, since it is off-chip.
The host has read and write access to the global memory, which allows this memory to be used to provide the input for, and read the output of, a kernel execution.

Threads are executed using the SIMT model. This means that each thread is executed independently with its own instruction address and local state (stored in its registers), but execution is organised in groups of 32 threads, called {\em warps}.
The threads in a warp execute instructions in lock-step, i.e.\ they share a program counter.
If the memory accesses of threads in a warp can be grouped together physically, i.e.\ if the accesses are coalesced, then the data can be obtained using a single fetch, which greatly improves the bandwidth compared to fetching physically separate data. 

\section{A GPU algorithm for term rewriting}

In this section, we address how a GPU can perform inner-most term rewriting to get the terms of a given TRS in normal form. Due to the different strengths and weaknesses of GPUs compared to CPUs, this poses two main challenges:
\begin{enumerate}
\item On a GPU, many threads (in the order of thousands) should be able to contribute to the computation;
\item GPUs are not very suitable for recursive algorithms. It is strongly advised to avoid recursion because each thread maintains its own stack requiring a large amount of stack space that needs to be allocated in slow global memory.
\end{enumerate}

We decided to develop a so-called \emph{topology-driven} algorithm~\cite{datadriven}, as opposed to a data-driven one. Unlike for CPUs, topology-driven algorithms are often developed for GPUs, in particular for irregular programs with complex data structures such as trees and graphs. In a topology-driven GPU algorithm, each GPU thread is assigned a particular data element, such as a graph node, and all threads repeatedly apply the same operator on their respective element. This is done until a fix-point has been reached, i.e., no thread can transform its element anymore using the operator. In many iterations of the computation, it is expected that the majority of threads will not be able to apply the operator, but on a GPU this is counterbalanced by the fact that many threads are running, making it relatively fast to check all elements in each iteration. In contrast, in a data-driven algorithm, typically used for CPUs, the elements that need processing are repeatedly collected in a queue before the operator is applied on them. Although this avoids checking all elements repeatedly, on a GPU, having thousands of threads together maintaining such a queue is typically a major source for memory contention.

In our algorithm, each thread is assigned a term, or more specifically a location where a term may be stored. As derivations are applied on a TRS, new terms may be created and some terms may be deleted. The algorithm needs to account for the number of terms dynamically changing between iterations.

First, we discuss how TRSs are represented on a GPU. Typically, GPU data structures, such as matrices and graphs, are array-based, and we also store a TRS in a collection of arrays. Each term is associated with a unique index $i$, and each of its attributes can be retrieved by accessing the $i$-th element of one of the arrays. This encourages coalesced memory access for improved bandwidth: when all threads need to retrieve the head symbol of their term, for instance, they will access consecutive elements of the array that stores head symbols. 
We introduce the following GPU data structures that reside in global memory:
\begin{itemize}
\item Boolean arrays \nfarray and \nffreeze keep track of which terms are in normal form, the first is used for writing and the second for reading;
\item Integer variable \n provides the current number of terms;
\item Array $\heads$ stores the head symbols of all the terms; 
\item Constant \maxarity refers to the highest arity among the function symbols in $F$;
\item Arrays $\argu_0, \ldots, \argu_{\maxarity-1}$ store the indices of the subterms of each term. Index 0 is never used. If $\argu_j[i] = 0$, for some $0 \leq j < {\maxarity-1}$, then the term stored at index $i$ has arity $j-1$, and all elements $\arg_j[i], \ldots, \arg_{\maxarity-1}[i]$ should be ignored.
\item Boolean flag \done indicates whether more rewriting iterations are needed;
\item Integer arrays $\refcounts$, $\refcountsread$ are used to write and read the number of references to each term, respectively. When a term is not referenced, it can be deleted.
\end{itemize}

Since GPUs have relatively little memory, some form of garbage collection is necessary to be able to reuse memory occupied by deleted terms. For this reason, we have the following additional data structures:
\begin{itemize}
\item Boolean flag \garbagecollecting indicates whether garbage collecting is needed;
\item Integer array \freeindices stores indices that can be reused for new terms;
\item Integer variables \nextfreebegin, \nextfreeend provide indices to remove elements from the front of \freeindices and add elements at the end, respectively;
\item Integer variable \nextfresh provides a new index, greater than the largest index currently occupied by a term in the term arrays. There, a new term can be inserted.
\end{itemize}

\begin{lstlisting}[language=pseudo,caption={The main loop of the rewrite algorithm, executed by the CPU},label=list:gpumain,captionpos=t,float=t,abovecaptionskip=-\medskipamount]
h_done = false;
while (!h_done) {
  done (*$\leftarrow$*) h_done;
  numBlocks = n / blockSize;
  refcounts_read = refcounts;
  nf_read = nf;
  derive<<numBlocks, blockSize>>(nf, nf_read, hss, arg(*$_0$*), (*$\ldots$*));
  h_next_fresh (*$\leftarrow$*) next_fresh;
  if (h_next_fresh > 0) {
    n = n + h_next_fresh;
    next_fresh (*$\leftarrow$*) 0;
  }
  h_done (*$\leftarrow$*) done;
  h_garbage_collecting (*$\leftarrow$*) garbage-collecting;
  if (h_garbage_collecting) {
    collect_free_indices <<numBlocks, blockSize>>((*$\ldots$*));
    garbage_collecting (*$\leftarrow$*) false;
  }
}
\end{lstlisting}

Listing~\ref{list:gpumain} presents the main loop of the algorithm, which is executed by the CPU. In it, two GPU kernels are repeatedly called until a fix-point has been reached, indicated by \done. To keep track of the progress, there are CPU counterparts of several variables, labeled with the `\texttt{h\_}' prefix. Copying data between CPU and GPU memory is represented by $\leftarrow$.

While the rewriting is not finished (line 2), the GPU \done flag is set to \textbf{false} (line 3), after which the number of thread blocks is determined. As the number of threads should be equal to the current number of terms, \n is divided by the preset number of threads per block (\texttt{blockSize}). After that, \refcounts is copied to \refcountsread, and \nfarray to \nffreeze. The reading and writing of the reference counters and normal form state is separated by the use of two arrays, to avoid newly created terms already being rewritten before they have been completely stored in memory.
The \texttt{derive} kernel is then launched for the selected number of blocks (line 7). This kernel, shown in Listing~\ref{list:gpuderive}, is discussed later. In the kernel, the GPU threads perform one rewrite iteration. Then, at lines 8-12, \n is updated in case the number of terms has increased. The \nextfresh variable is used to count the number of new terms placed at fresh indices, i.e., indices larger than \n when \texttt{derive} was launched.

Finally, with \garbagecollecting, it is monitored whether some indices of deleted terms need to be gathered in the \freeindices list. This gathering is done by the \texttt{collect\_free\_indices} kernel: if a thread detects that the reference counter of its term is $0$, it decrements the counters of the subterms and the index to the term is added to the \freeindices list. Atomic memory accesses are used to synchronise this. Notice that \freeindices is in device memory and no unnecessary data is transferred back and forth between host and device.


\begin{lstlisting}[language=pseudo,caption={The derive kernel, executed by a GPU thread},label=list:gpuderive,captionpos=t,float=t,abovecaptionskip=-\medskipamount]
derive (...) {
  if (tid >= (*$n$*)) { return; }
  refcount = refcounts_read[tid];
  if (refcount > 0) {
    start_rewriting = !nf_read[tid];
    if (start_rewriting) {
      if ((*\textit{all subterms are in normal form}*)) {
        switch (hss[tid]) {
          case (*$f$*):  rewrite(*$_f$*)((*$\ldots$*));
          (*$\ldots$*)
          default: nf[tid] = true;
        }
      }
      done = false;
    }
  }
  else {
    garbage_collecting = true;
  }
}
\end{lstlisting}

In Listing~\ref{list:gpuderive}, the GPU \texttt{derive} kernel is described. When the kernel is launched for \texttt{numBlocks}$\cdot$\texttt{blockSize} threads, each of those threads executes the kernel to process its term. The global ID of each thread is \texttt{tid}. Some threads may not actually have a term to look at (if \n is not divisible by \texttt{blockSize}), therefore they first check whether there is a corresponding term (line 2). If so, the value of the reference counter for the term is read (line 3), and if it is non-zero, a check for rewriting is required. Rewriting is needed if the term is not in normal form (line 5) and if all its subterms are in normal form. The latter condition is briefly referred to at line 7. To avoid repetitive checking of subterms in each execution of the \texttt{derive} kernel, every thread keeps track of the last subterm it checked in the previous iteration. If rewriting is required, the suitable \texttt{rewrite}$_f$ function is called, depending on the head symbol of the term (lines 8-12). If no function is applicable, the term is in normal form (line 11). Finally, \done is set to \textbf{false} to indicate that another rewrite iteration is required. Alternatively, if the reference counter is $0$, the \garbagecollecting flag is set. This causes the \texttt{collect\_free\_indices} kernel to be launched after the \texttt{derive} kernel (see Listing~\ref{list:gpumain}).

\begin{lstlisting}[language=pseudo,caption={An example rewrite function for the rule \texttt{Plus(Zero,X)$\to$\texttt{X}}, executed by a GPU thread},label=list:gpurewritef,captionpos=t,float=t,abovecaptionskip=-\medskipamount]
rewrite(*$_\texttt{Plus}$*)((*$\ldots$*)) {
  r_0 = arg0[tid];
  r_hs_0 = hss[r_0];
  if (r_hs_0 == Zero) {
    r_1 = arg1[tid];
    r_hs_1 = hss[r_1];
    hss[tid] = r_hs_1;
    copy_term_args(refscount, arg0, arg1, r_1, tid, r_hs_1);
    atomicSub(&refcounts[r_0], 1);
    atomicSub(&refcounts[r_1], 1);
    nf[tid] = true;
    return;
  }
  (*...\textit{Check applicability of other Plus-rules}*)
}
\end{lstlisting}

Given a TRS, the \texttt{rewrite}$_f$ functions are automatically generated by a code generator we developed, to directly encode the rewriting in CUDA code. Listing~\ref{list:gpurewritef} provides example code for the rewrite rule \texttt{Plus(Zero,X)}$\rightarrow$\texttt{X}, which expresses that adding $0$ to some number \texttt{X} results in \texttt{X}. Applicability of this rule is checked by the \texttt{rewrite}$_\texttt{Plus}$ function, which may also involve other rules for terms with head symbol \texttt{Plus}. First, to check applicability, the index of the first subterm is retrieved, and with it, the head symbol of that term (lines 2-3). If the head symbol is \texttt{Zero} (line 4), the index to the second subterm is retrieved (line 5). The rewriting procedure should ensure that the term at position \texttt{tid} is replaced by \texttt{X}.

When constructing terms, sharing of subterms is applied whenever possible. For instance, if a term \texttt{F(X,X)} needs to be created, the index to \texttt{X} would be used twice in the new term, to make sure both subterm entries point to the same term in physical memory. When rewriting the term itself, however, as in the example, we have to copy the attributes of \texttt{X} to the location \texttt{tid} of the various arrays, to ensure that all terms referencing term \texttt{tid} are correctly updated. 

This copying of terms is done by first copying the head symbol (lines 6-7), and then the indices of the subterms, which is done at line 8 by the function \texttt{copy\_term\_args}; it copies the number of subterms relevant for a term with the given head symbol, and increments the reference counters of those subterms. Next, the reference counters of \texttt{Zero} and \texttt{X} are atomically decremented (since the term \texttt{Plus(Zero, X)} is removed) (lines 9-10), and we know that the resulting term is in normal form, since \texttt{X} is in normal form (line 11).

\begin{lstlisting}[language=pseudo,caption={The get\_new\_index GPU function, executed by a GPU thread},label=list:gpugetnewindex,captionpos=t,float=t,abovecaptionskip=-\medskipamount]
get_new_index((*$\ldots$*)) {
  if (tid >= (*$n$*)) { return; }
  n_begin = next_free_begin; n_end = next_free_end;
  new_id = 0;
  if (n_begin < n_end) {
    n_begin = atomicInc(&next_free_begin);
    if (n_begin < n_end) {
      new_id = free_indices[n_begin];
    }
  }
  if (new_id == 0) {
    new_id = atomicInc(&next_fresh) + n;
  }
  return new_id;
}
\end{lstlisting}

Finally, we show how new indices are retrieved whenever a new term needs to be created. In the example of Listing~\ref{list:gpurewritef}, this is not needed, as the RHS of the rule has no new subterms, but for a rule such as \texttt{Plus(S(0), X)}$\rightarrow$\texttt{S(X)}, with \texttt{S} representing the successor function (i.e., \texttt{S(0)} represents $1$) a new term \texttt{S(X)} needs to be created, with its only subterm entry pointing to the term referenced by the second subterm entry of the LHS.

Listing~\ref{list:gpugetnewindex} shows how we retrieve a new index.
Due to garbage collection, a number of indices may be available in the first \n entries of the input arrays which are currently used. These are stored in the \freeindices array, from index \nextfreebegin to index \nextfreeend. If this array is not empty (line 5), \nextfreebegin is atomically incremented to claim the next index in the \freeindices array (line 6). If this increment was not performed too late (other threads have not since claimed all available indices), the index is stored in \texttt{new\_id} (lines 7-9). Otherwise, a new index must be added at the end of the current list of terms. The variable \nextfresh is used for this purpose: $\nextfresh + \n$ can be used as a new index, and \nextfresh needs to be incremented for use by another thread.

\section{Evaluation}




\begin{table}[t]
  \caption{Comparison of the CPU and GPU.}
  \label{table:comparison-cpu-gpu}
\scalebox{0.8}{
  \begin{tabular}{lllrrr}
    Type & Year & Name & Mem (GB) & BW aligned (GiB/s) & BW random (GiB/s) \\
    \hline
    CPU & 2017 & Intel Core i5-7600 & 32 & 25.7 & 0.607 \\
    GPU & 2018 & NVIDIA Titan RTX & 24 & 555 & 22.8 \\ 
\end{tabular}
}
\end{table}

In this section we provide insight into the performance of the GPU rewriter.
We do this in two ways: We compare our GPU rewriter with a sequential recursive left-most inner-most
rewriter for the CPU (1) and (2) we analyze to what extent we make good use of 
the GPU resources.  Because CPUs and GPUs differ widely in architecture, it is often subject of debate whether a comparison is fair \cite{Lee2010Debunking}.  We therefore include the second way of evaluating.

Table~\ref{table:comparison-cpu-gpu} shows a comparison of the used CPU and GPU.  CPUs are optimized for latency: finish the program as soon as possible.  To achieve this, they have  deep cache hierarchies, deep execution pipelines and dedicated logic for branch prediction and extracting parallelism from a serial instruction stream.  In contrast, GPUs are optimized for throughput: process as many elements per time unit as possible.  For GPUs, parallelism is explicit, one instruction is issued for multiple threads, and the architecture is specifically designed to hide memory latency times by scheduling new warps immediately after a memory access.  The differences between architectures are highlighted by the last two columns that show that the bandwidth of the GPU for aligned access is vastly superior to that of the CPU.  Even the bandwidth for random accesses on the GPU almost reaches the bandwidth for aligned accesses on the CPU.

We measure the performance of the CPU and GPU rewriter in \emph{rewritten terms per second}.  Given a TRS, both the GPU and the CPU rewriter are generated by a \texttt{Python 2.7} script. The script uses \texttt{TextX}~\cite{Dejanovic2017} and \texttt{Jinja 2.11}\footnote{\url{https://jinja.palletsprojects.com}.} to parse a TRS and generate a \texttt{rewrite}$_f$ function for every rewritable head symbol $f$ in the TRS. The code generated for the GPU rewriter is CUDA C++ with CUDA platform 10.1.  For the CPU rewriter the code generated is in C++.  The same \texttt{rewrite}$_f$ functions are used, and thus the rewrite rules are exactly the same and the CPU and GPU implementations rewrite exactly the same number of terms.

We evaluate the GPU rewriter with a TRS for sorting one or more lists of Peano numbers with merge sort (see the example in Section~\ref{sec:prelim}) and with a TRS that transforms a large number of terms.\footnote{See Appendix~\ref{sec:appendix} for a detailed description of the two TRSs.}  These TRSs accentuate the capabilities of the GPU and the CPU.  Merge sort is a divide-and-conquer algorithm amenable to parallelism, but splitting up and combining lists are highly sequential operations. 

\begin{figure}[t]

    \centering
    \scalebox{0.75}{
        \begin{subfigure}[b]{0.495\textwidth}
        \includegraphics[width=\textwidth]{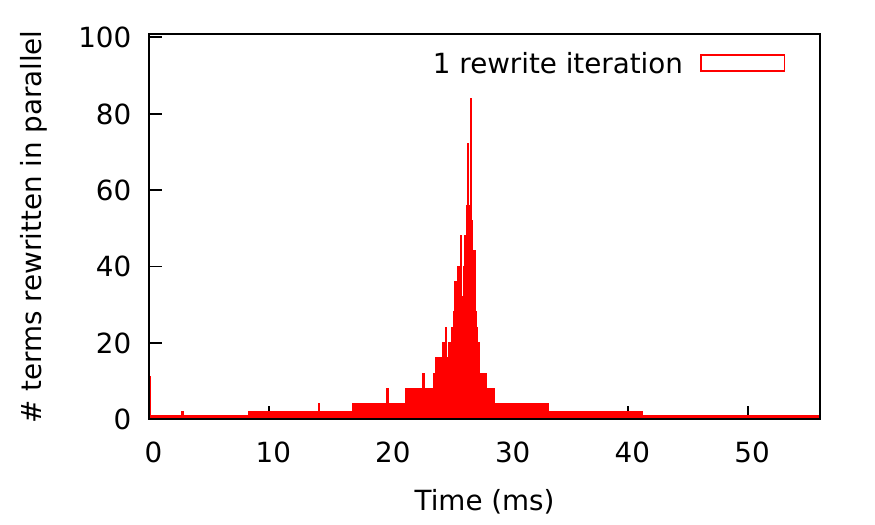}
        \caption{Parallelism across execution time}
        \label{fig:mergesort-50-parallelism}
    \end{subfigure}
    \begin{subfigure}[b]{0.495\textwidth}
    \includegraphics[width=\textwidth]{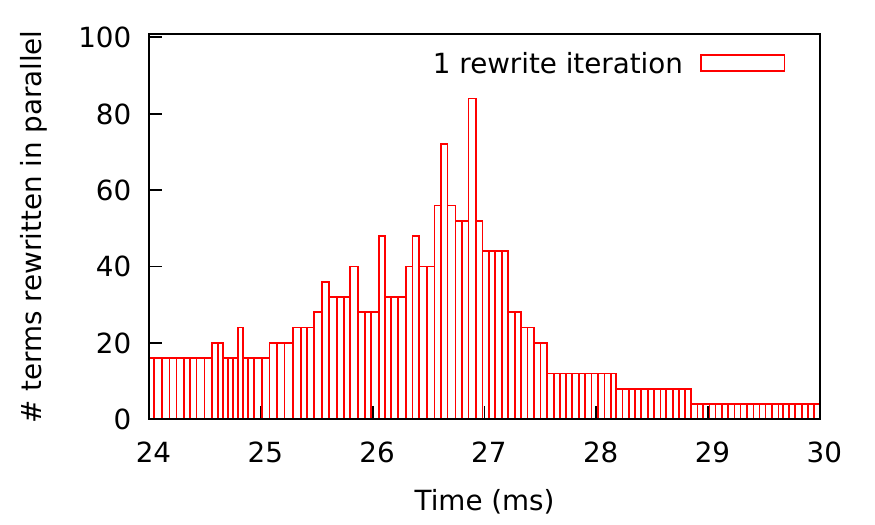}
        \caption{Parallelism zoomed in}
        \label{fig:mergesort-50-parallelism-zoomed}
    \end{subfigure}
    }
 	\caption{Merge sort of 50 elements on the GPU.}
	\label{fig:mergesort-50}
\end{figure}

Figure~\ref{fig:mergesort-50} shows a merge sort performed on a single list of 50 elements.  The width of a red box (too small for Fig.~\ref{fig:mergesort-50-parallelism}, see the zoomed in version in Fig.~\ref{fig:mergesort-50-parallelism-zoomed}) represents the time of a GPU rewrite step (the \texttt{derive} statement on line 7 in Listing~\ref{list:gpumain}) whereas the height represents how many terms are rewritten in parallel in this rewrite step.  The figure shows that there are long tails of a low degree of parallelism before and after a brief peak of parallelism.  Given Amdahl's law that states that speedup is severely limited with a low degree of parallelism \cite{Amdahl1967Validity}, it is clear that merge sort on single lists is not parallel enough for GPUs.

The performance of the GPU of $74\times10^3$ terms/s versus the CPU $97\times10^6$ terms/s highlights a different issue, namely Gustafson's Law~\cite{Gustafson1988Reevaluating}:  To overcome the overhead of using a highly parallel machine, we need a large problem with a high degree of parallelism to highlight the capabilities of the GPU.
In order to benchmark this potential, we use the merge sort TRS applied on multiple lists: The Tree merge sort is given by a binary tree with a list of numbers at every leaf. All these lists are sorted concurrently using the same merge sort as in the previous example. The parallelism is exponential w.r.t.\ the depth of the tree. 

\begin{table}[t]
  \caption{Performance of the rewrite systems.}
  \label{table:performance}
\scalebox{0.8}{
  \begin{tabular}{lllrrr}
    Application & CPU rewritten terms/s & GPU rewritten terms/s & Speedup \\
    \hline
    Merge sort 50 & $97\times10^6$ & $74\times10^3$ & $0.76\times10^{-3}$\\
    Tree merge sort 23 5 & $113\times10^6$ & $387\times10^6$ & $3.34$ \\
    Transformation tree 22 & $26.5\times10^6$ & $3.12\times10^9$ & $11.7$ \\
\end{tabular}
}
\end{table}

Table~\ref{table:performance} shows that the GPU outperforms the CPU more than a factor of three for a binary tree of 23 levels deep of merge sorts of lists of 5 numbers, which translates to sorting approximately 8 million lists.  Finally, to understand the true potential of the GPU, we designed the Transformation tree benchmark that expands a binary tree to 22 levels deep (so 4 million leaves) where each leaf is rewritten 26 times.  On this benchmark, the GPU rewriter is more than a factor 10 faster than the CPU rewriter, achieving 3.12 billion rewrites per second on average over the complete execution time, but sustaining around 6 billion rewrites per second for half of the execution time (the rest is setting/breaking down the tree).

\begin{figure}[t]
    \centering
    \scalebox{0.75}{
    \begin{subfigure}[b]{0.495\textwidth}
        \includegraphics[width=\textwidth]{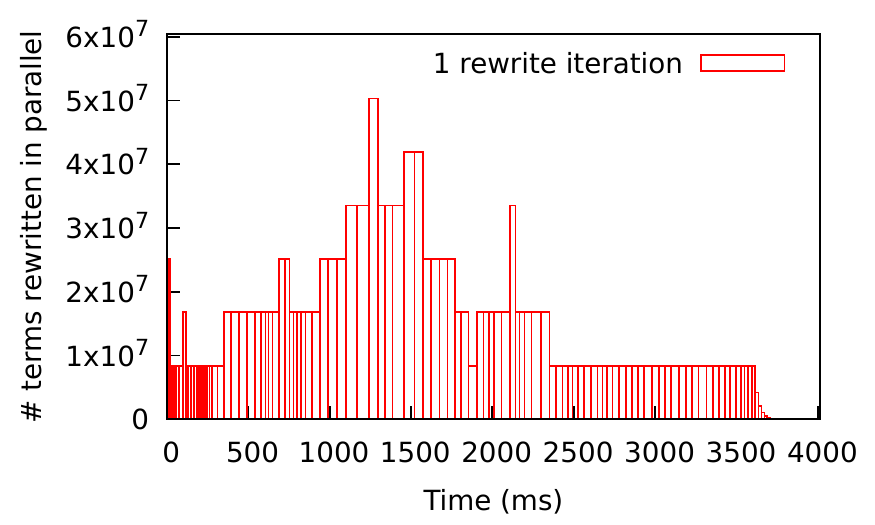}
        \caption{Parallelism}
        \label{fig:mergesort-tree-parallelism}
    \end{subfigure}
    \begin{subfigure}[b]{0.495\textwidth}
    \includegraphics[width=\textwidth]{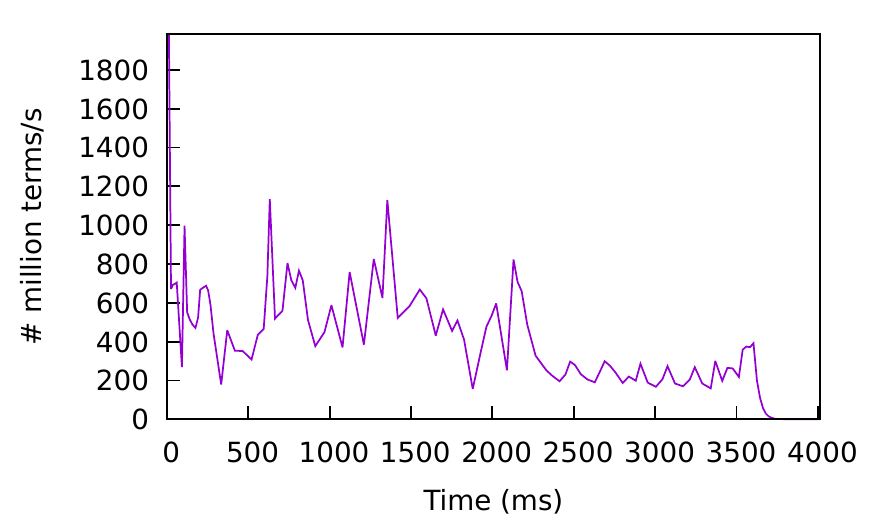}
        \caption{Throughput}
        \label{fig:mergesort-tree-throughput}
    \end{subfigure}
    }
    \scalebox{0.75}{
    \begin{subfigure}[b]{0.495\textwidth}
        \includegraphics[width=\textwidth]{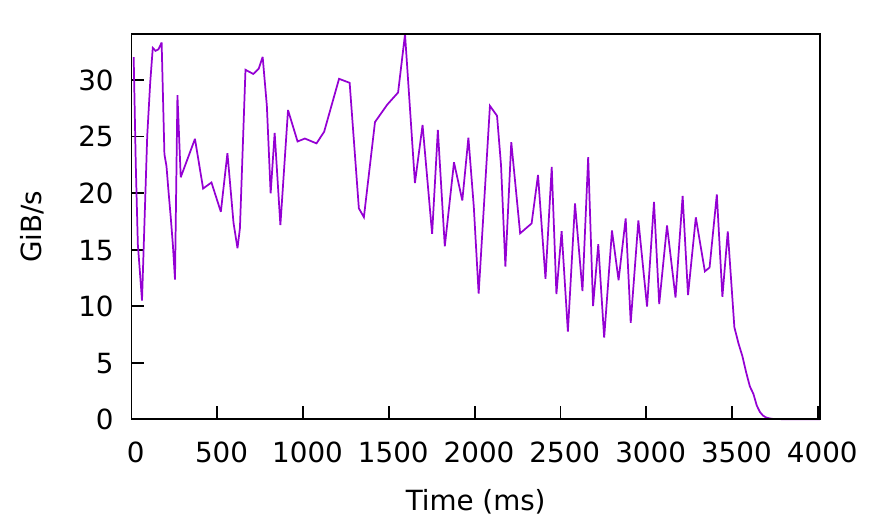}
        \caption{Bandwidth random access}
        \label{fig:mergesort-tree-bw-random}
    \end{subfigure}
    \begin{subfigure}[b]{0.495\textwidth}
        \includegraphics[width=\textwidth]{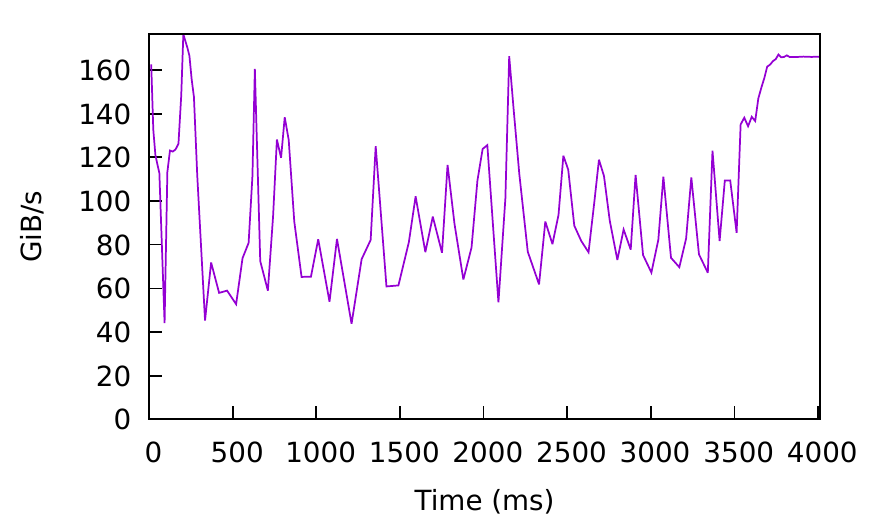}
        \caption{Bandwidth coalesced access}
        \label{fig:mergesort-tree-bw-coalesced}
    \end{subfigure}
    }	   
 	\caption{Merge sort of a tree of 23 deep with lists of 5 elements.}
	\label{fig:mergesort-tree}
\end{figure}

To understand the performance better we focus on the more realistic Tree merge sort benchmark.
Figure~\ref{fig:mergesort-tree} shows several graphs for the execution with which we can analyze the performance.  Figure~\ref{fig:mergesort-tree-parallelism} shows that this rewrite system shows a high degree of parallelism for almost all of the execution time.  Figure~\ref{fig:mergesort-tree-throughput} shows that the rewriter shows a high throughput with peaks up to 1 billion terms rewritten per second.

The bottom two figures highlight to what extent we use the capabilities of the GPU.  Usually, the performance of a GPU is measured in GFLOPS, floating point operations per second, for compute intensive applications or GiB/s for data intensive applications.  Since term rewriting is a symbolic manipulation that does not involve any arithmetic, it is data intensive.  From Table~\ref{table:comparison-cpu-gpu} we have seen that the maximum bandwidth our GPU can achieve is 555~GiB/s for aligned accesses and 22.8~GiB/s for random accesses.  Since term rewriting is an irregular problem with a high degree of random access (to subterms that can be anywhere in memory), we focus on the bandwidth for random accesses.  Table~\ref{table:tree-mergesort-bw} shows that the overall random access bandwidth of the GPU implementation reaches 18.1~GiB/s which is close to the benchmarked bandwidth.  In addition, the aligned bandwidth of 95.7~GiB/s confirms that term rewriting is indeed an irregular problem and that aligned bandwidth is less of a bottleneck.  The bottom two graphs in Figure~\ref{fig:mergesort-tree} show the measured bandwidth over time.

\begin{table}[t]
  \caption{Performance the tree merge sort in terms of bandwidth.}
  \label{table:tree-mergesort-bw}
\scalebox{0.8}{
  \begin{tabular}{lrrr}
    Application & BW random (GiB/s) & BW aligned (GiB/s)\\
    \hline
    Tree merge sort 23 5 & 18.1 & 95.7 \\ 
\end{tabular}
}
\end{table}

Although we are close to the random access bandwidth of the GPU, this does not mean that we have reached the limits of term rewriting on GPUs.  It does mean however, that to achieve higher performance with term rewriting on GPUs, it is necessary to introduce more regularity into the implementation, reducing the random memory accesses. It also means that other often used strategies to improve graph algorithm, like reducing branch divergence will probably not yield significant performance increase. In addition, the results we present clearly show the different capabilities of GPUs and CPUs. An interesting direction for future work is to create a hybrid rewrite implementation that can switch to a GPU implementation when a high degree of parallelism is available.

\bibliography{references}

\appendix
\section{TRSs used as benchmarks}
\label{sec:appendix}

\paragraph*{TRS for Tree merge sort of multiple lists}\label{appendix:multisort_trs}

Listing~\ref{list:msortcomplete} gives the concrete TRS description used in our benchmarks of the merge sorting on lists. Natural numbers, lists and trees are given by the inductive types \texttt{Nat} , \texttt{List}, and \texttt{Tree}, respectively. The Booleans are given by the two constant functions \rwrule{True()} and \rwrule{False()}, and the two functions on natural numbers, \rwrule{Lt} and \rwrule{Gt}, evaluate like the less than and greater than function, respectively. 

The functions \rwrule{Even(L)} and \rwrule{Odd(L)} with \rwrule{L} a subterm of type \texttt{List} will evaluate to a \texttt{List} that contains only the even and odd elements of \texttt{L}, respectively. 

The essential functions that perform the merge sort are \rwrule{Merge}, \rwrule{Merge2} and \rwrule{Sort}, \rwrule{Sort2}. The \rwrule{Merge} function evaluates to a term representing the sorted merged list of the two subterms. The \rwrule{Sort} function represents the implementation of merge sort and splits the list subterm until it consists of at most one term.  The \rwrule{Sort2} function is where the parallelism of merge sort can be observed. In the RHS of this rule the two subterms of \rwrule{Merge}, consisting of the recursive call of \rwrule{Sort} can be evaluated in parallel for both the even and odd partition of the list \texttt{L}.

To the basic TRS for merge sorting a list, we have added a tree data structure, to support input terms consisting of multiple lists. The TRS rewrites all the \rwrule{Sort} terms in the leaves of a given tree of lists in parallel. While a list of lists would result in a sequential evaluation of the lists, a tree structure allows parallel evaluation. 

\begin{lstlisting}[caption={TRS for mergesort on multiple lists}, label={list:msortcomplete}]
sort  Nat  = struct Zero() | S(Nat) | Len(List);
      Bool = struct True() | False() | Lt(Nat, Nat) | Gt(Nat, Nat);
      List = struct Nil() | Cons(Nat, List) | Merge(List, List) |
                    Merge2(Bool, Nat, List, Nat, List) | Even(List) |
                    Odd(List) | Sort(List) | Sort2(Bool, List);
      Tree = struct Leaf(List) | Node(Tree,Tree);

var X : Nat; Y : Nat; B : Bool; L : List; M : List;

eqn	
  Len(Nil()) = Zero();
  Len(Cons(X, L)) = S(Len(L));

  Merge(Nil(), M) = M;
  Merge(L, Nil()) = L;
  Merge(Cons(X, L), Cons(Y, M)) = Merge2(Lt(X,Y), X, L, Y, M);

  Merge2(True(), X, L, Y, M) = Cons(X, Merge(L, Cons(Y, M)));
  Merge2(False(), X, L, Y, M) = Cons(Y, Merge(Cons(X, L), M));

  Sort(L) = Sort2(Gt(Len(L), S(Zero())), L);
  Sort2(False(), L) = L;
  Sort2(True(), L) = Merge(Sort(Even(L)), Sort(Odd(L)));

  Even(Nil()) = Nil();
  Even(Cons(X, L)) = Cons(X, Odd(L));
  Odd(Nil()) = Nil();
  Odd(Cons(X, L)) = Even(L);

  Gt(Zero(), Zero()) = False();
  Gt(Zero(), S(Y)) = False();
  Gt(S(X), Zero()) = True();
  Gt(S(X), S(Y)) = Gt(X, Y);

  Lt(Zero(), Zero()) = False();
  Lt(Zero(), S(Y)) = True();
  Lt(S(X), Zero()) = False();
  Lt(S(X), S(Y)) = Lt(X, Y);
  
Input Node(Node(Leaf(Sort(Cons(S(Zero()),$\dots$)), Leaf(Sort($\dots$))),Node($\dots$));
\end{lstlisting}
\paragraph*{TRS for Transformation tree}\label{appendix:transformation_tree}
Listing~\ref{list:dummytree} gives the TRS which is constructed to showcase massive parallelism. The \texttt{Tree} type consists of nodes and leaves in the form of symbols \texttt{A} through \texttt{Z} and \texttt{End}. A tree of depth $N$ with \texttt{A}'s in the leaves is generated by rewriting the term \rwrule{Expand(N)}. The symbol \texttt{A} is rewritten to a \texttt{B} and this is rewritten again until \texttt{End} is reached. This is done for all leaves, and can be done in one single massively parallel rewrite step. 

The two rules \rwrule{Expand} and \rwrule{Expand2} are technically the same function but are used to prevent the subterms from being equal. If the subterms where equal the tree generated would have only a single term representing the leaves, due to sharing, and there would be no parallelism left.

\begin{lstlisting}[caption={Transformation tree of depth 22}, label={list:dummytree}]
sort Nat = struct Zero() | Suc(Nat);
     Tree = struct A() | B() | $\dots$ | Z() | End() |
              Node(Tree,Tree) | Expand(Nat) | Expand2(Nat);

var O : Tree;
    P : Tree;
    X : Nat;

eqn Expand(Zero()) = A();
    Expand(S(X)) = Node(Expand(X),Expand2(X));

    Expand2(Zero()) = A();
    Expand2(S(X)) = Node(Expand(X), Expand2(X));

    A() = B();
    B() = C();
    C() = D();
    D() = E();
    $\dots$
    Z() = End();

Input Expand(Suc(Suc(Suc($\dots$Suc(
	     Zero()
	)$\dots$))));
\end{lstlisting}

\end{document}